 \documentstyle[eqsecnum,prb,aps]{revtex}

\begin{document}

%%%%%%%%%%%%%%%%%%%%%%%%%%%%%%%%%%%%%%%%%%%%%%%%%%%%%%%%%%%

\draft
\preprint{}
\title{Local density of states in the vortex lattice 
in a type II superconductor}
\author{Masanori Ichioka, Nobuhiko Hayashi, and Kazushige Machida}
\address{Department of Physics, Okayama University, Okayama 700, Japan}
\date{
%\today
October 15, 1996}
\maketitle
\begin{abstract}
Local density of states (LDOS) in the triangular vortex lattice is  
investigated based on the quasi-classical Eilenberger theory. 
We consider the case of an isotropic $s$-wave superconductor with 
the material parameter appropriate to ${\rm NbSe_2}$. 
At a weak magnetic field, the spatial variation of the LDOS shows 
cylindrical structure around a vortex core. 
On the other hand, at a high field where the core regions substantially 
overlap each other, the LDOS is sixfold star-shaped structure due to 
the vortex lattice effect.  
The orientation of the star coincides with the experimental data of the 
scanning tunneling microscopy. 
That is, the ray of the star extends toward the nearest-neighbor 
(next nearest-neighbor) vortex direction at higher (lower) energy. 
\end{abstract}
\pacs{PACS numbers: 74.60.Ec, 74.25.Jb, 61.16.Ch }

\widetext
%\narrowtext

%%%% s1 %%%%%%%%%%%%%%%%%%%%%%%%%%%%%%%%%%%%%%%%%%%%%%%%%%%%%%%
\section{Introduction}
\label{sec:1}

Since the success of observing the vortex core image by Hess {\it et al.} in 
the scanning tunneling microscopy (STM) experiments, 
it has been possible to investigate  experimentally the local density 
of states (LDOS) around a vortex core in type II 
superconductors.\cite{Hess1,Hess2,Hess3,Hess4,Hess5,Renner} 
This investigation has revealed a rich internal electronic structure 
associated with a vortex core, and should be important for understanding 
the conventional and unconventional superconductors. 
Then, to extract further information from the experimental data, 
the detailed calculation of the LDOS is expected from the theoretical side. 

Hess {\it et al.} have done a series of beautiful STM 
experiments on a layered hexagonal compound 2H-${\rm NbSe_2}$ ($T_c$=7.3K) 
to reveal the detailed spatially resolved electronic structure around a 
vortex core.\cite{Hess1,Hess2,Hess3,Hess4,Hess5}  
They provided direct images of individual vortices and the flux line lattice.
Their experiment is the first one detecting the quasiparticle 
state bound to the vortex core, which was theoretically predicted 
by Caroli, de Gennes and Matricon.\cite{Caroli}  
Further, the remarkable result is the sixfold star-shaped LDOS around a 
vortex.\cite{Hess2,Hess3,Hess4,Hess5} 
Their results are summarized as follows when the magnetic field 
is applied perpendicular to the hexagonal plane: 
(1) The LDOS for quasiparticle excitations has a sixfold star-shape 
centered at a core. 
(2) The orientation of this star depends on the bias energy. 
At zero bias, the ``ray'' of the star extends away from the 
nearest-neighbor direction where the conventional $60^\circ$ Abrikosov 
vortex lattice is formed. 
Upon increasing the bias voltage, the star rotates by $30^\circ$ and the 
ray extends to the  nearest-neighbor direction. 
(See Fig. 4 in Ref. \onlinecite{Hess2} or Fig. 1 in Ref. \onlinecite{Hess4}.) 
(3)  In the intermediate bias voltage, a ray splits into a pair of nearly 
parallel rays, keeping its direction fixed (see the STM image for 0.48 mV in 
Fig. 1 of Ref. \onlinecite{Hess4}).  

Recently, on the other hand, Maggio-Aprile {\it et al.} have succeeded 
in observing the STM image of the vortex cores on a high-$T_c$ 
superconductor ${\rm YBa_2Cu_3O_7}$.\cite{Maggio}  
One of the point in analyzing their STM image of vortex is how the gap 
anisotropy ($d_{x^2-y^2}$-wave in the high $T_c$ superconductor) affects 
the LDOS around a vortex. 

To understand these experimental results, the concrete form of the LDOS 
structure is expected to be calculated from the theoretical side. 
So far, the LDOS around a vortex was calculated by Gygi and 
Schl\"uter,\cite{Gygi1,Gygi2} and Shore {\it et al.}\cite{Shore} from the 
Bogoliubov-de Gennes (BdG) equation.  
The calculations of the LDOS based on the quasi-classical Eilenberger (QCE) 
theory\cite{Eilenberger} were done by Klein,\cite{Klein90} and Ullah 
{\it et al.}\cite{Ullah} for an isotropic $s$-wave superconductors. 
While these calculations showed that the vortex image observed by STM is due 
to the quasiparticle state bound to the vortex core, they investigated 
only the case of an isolated single vortex, and based on the assumption 
of the cylindrically symmetric vortex core structure. 
In a $d_{x^2-y^2}$-wave superconductor, the LDOS around an isolated single 
vortex was calculated by  Schopohl and Maki,\cite{Schopohl,Maki} and 
Ichioka {\it et al.}\cite{Ichioka} 
However, the LDOS in the vortex lattice case has not been calculated 
so far. 

On the other hand, the STM experiments associated with the vortex core are 
usually performed at high magnetic fields, where the distance between 
vortices is short and the overlap of the vortex core with that of the 
nearest-neighbor vortices can not be neglected. 
In this situation, the LDOS around a vortex core is expected to 
break cylindrical symmetry and show sixfold symmetric structure 
when the vortex lattice forms an triangular lattice. 
The purpose of this paper is to calculate the LDOS in the triangular 
vortex lattice by using the QCE theory, and clarify its sixfold symmetric 
structure.  
It is  expected to be observed at a high magnetic field. 
For the case cylindrical symmetry is broken, 
the QCE approach is more suitable than the BdG approach.

As for the sixfold symmetric structure of the LDOS observed by Hess 
{\it et al}.,\cite{Hess2,Hess3,Hess4,Hess5} 
Gygi and Schl\"uter \cite{Gygi2,Gygi3} discussed it on the basis of 
the BdG theory by introducing a sixfold symmetric perturbation term 
and using their numerical solutions of the cylindrical symmetric case.
While they explained the above-mentioned experimental features (1) and 
(2), the origin of their perturbation term is not clear, and it is uncertain 
whether the sixfold symmetric term can be treated by a perturbation theory. 
We note that their theory cannot determine the absolute orientation of 
the star relative to the vortex lattice configuration, since 
in their theory the orientation is determined by the sign of the 
perturbation term, which is given as an assumption.

As for the origin of the sixfold symmetric vortex structure, 
the following three possibilities are enumerated within the 
weak coupling BCS theory; the effect of the vortex lattice 
(that is, the effect of nearest-neighbor vortices), 
the effect of a sixfold symmetrically anisotropic $s$-wave pairing, 
and the effect of the sixfold symmetrically anisotropic density of states 
at the Fermi surface. 
In this paper, we focus on the effect of the vortex lattice by calculating 
the LDOS naively in the vortex lattice case.
The contribution of this effect should be clarified before considering 
other anisotropic effects. 
The contribution of the other anisotropic effects is investigated 
elsewhere.\cite{Hayashi}

The quasi-classical calculation in the vortex lattice case was so 
far performed by Klein in the case of low-$\kappa$ type II superconductors 
such as Ta or Nb.\cite{Klein87}  
He calculated the spatial variation of the pair potential and magnetic field, 
where the Eilenberger equation is solved self-consistently in the Matsubara 
frequency and he succeeded in solving it by either the so-called symmetry 
method or the so-called explosion method.\cite{Klein87,Thuneberg}  
He also tried the calculation of the LDOS,\cite{Klein89,Pottinger} 
where the Eilenberger equation is solved in real energy instead of 
the Matsubara frequency by using the self-consistently obtained 
pair potential and vector potential. 	
In this real energy case, he could calculate only the momentum-resolved 
LDOS for specific ${\bf k}$ (the relative momentum of the Cooper pair)  
direction because only the symmetry method was used. 
In this paper, we calculate the LDOS following the method suggested by 
Klein.\cite{Klein87,Klein89,Pottinger}
However, as we succeed in solving the Eilenberger equation in the real 
energy case for arbitrary ${\bf k}$ directions by using the explosion method, 
we can  calculate the momentum-resolved LDOS for arbitrary ${\bf k}$ 
directions. 
Then we can obtain the LDOS integrated over all ${\bf k}$ directions. 

The LDOS in the vortex lattice is an important physical quantity, 
since it can be observed directly by the STM experiments. 
And further, it can be a clue of estimating the 
transfer of the quasi-particle bound states between vortex cores. 
This transfer leads to the band structure of the bound states as suggested 
by Canel.\cite{Canel}  
If the enough transfer exists, it makes the de Haas-van Alphen (dHvA) 
oscillation possible even in the superconducting state, which is observed 
in the materials such as ${\rm NbSe_2}$\cite{Graebner} and 
${\rm YBa_2Cu_3O_7}$.\cite{Fowler}  

In our calculations, we consider the same situation 
as that Hess {\it et al.} performed their STM experiments on 
${\rm NbSe_2}$.\cite{Hess1,Hess2,Hess3,Hess4,Hess5} 
We use the material parameter appropriate to ${\rm NbSe_2}$, and 
assume that the Fermi surface is two dimensional, which is appropriate to 
${\rm NbSe_2}$.
The magnetic field is applied perpendicular to the hexagonal plane, 
that is, along the $z$ axis. 
The case of the triangular vortex lattice in an isotropic $s$-wave 
superconductor is considered in the clean limit. 
The Fermi surface and the energy gap of the superconductivity are assumed 
to be isotropic in order to exclude other origins of anisotropy and to 
clarify the vortex lattice effect on the LDOS around a vortex. 
Throughout the paper, energies and lengths are measured in units of the 
uniform gap $\Delta_0$ at $T=0$ and the coherence length 
$\xi_0=v_{\rm F}/\Delta_0 =\pi \xi_{\rm BCS}$ ($v_{\rm F}$ 
is the Fermi velocity and $\xi_{\rm BCS}$ BCS coherence length), respectively. 
The magnetic field and the vector potential are, respectively, scaled 
by $\phi_0/\xi^2$ and $\phi_0/\xi$, where $\phi_0$ is the flux quantum. 

The rest of this paper is organized as follows. 
    In Sec. \ref{sec:2}, we describe the method of calculation based
on the QCE theory.
    Section \ref{sec:3} contains numerically obtained results about  
the LDOS of the vortex lattice.
    The summary and discussions are given in Sec. \ref{sec:4}.

%%%% s2 %%%%%%%%%%%%%%%%%%%%%%%%%%%%%%%%%%%%%%%%%%%%%%%%%%%%%%%
\section{Quasi-classical Eilenberger theory}
\label{sec:2}

The calculation based on the Eilenberger theory is performed by following 
the method suggested by Klein.\cite{Klein87,Klein89,Pottinger} 
First, we obtain the pair potential and vector potential self-consistently 
by solving the Eilenberger equation in the Matsubara frequency. 
Next, using them, we calculate the LDOS by solving the Eilenberger equation 
in the real energy instead of the Matsubara frequency. 

In our calculation, the unit vectors of the vortex lattice are given by 
${\bf r}_1=(a_x,0)$, ${\bf r}_2=(\zeta a_x,a_y)$. 
As we consider a triangular lattice, we set $a_y/a_x=\sqrt{3}/2$ and 
$\zeta=1/2$. 
The microscopic magnetic field ${\bf H}({\bf r})=(0,0,H({\bf r}))$ is 
divided into an external field $\bar {\bf H}=(0,0,\bar H)$ and an 
internal field ${\bf h}({\bf r})=(0,0,h({\bf r}))$, 
%%%
\begin{equation}
{\bf H}({\bf r})=\nabla\times {\bf A}({\bf r}) 
=\bar {\bf H} + {\bf h}({\bf r}), 
\label{eq:2.1}
\end{equation}
%%%
where the spatial average of ${\bf h}({\bf r})$ vanishes. 
Therefore, the vector potential ${\bf A}({\bf r})$ is also divided 
into two parts, 
%%%
\begin{equation}
{\bf A}({\bf r})=\frac{1}{2}\bar {\bf H} \times {\bf r} + {\bf a}({\bf r})
\label{eq:2.2}
\end{equation}
%%%
in the symmetric gauge.  
From Eqs. (\ref{eq:2.1}) and (\ref{eq:2.2}), we obtain 
${\bf h}({\bf r})=\nabla\times {\bf a}({\bf r})$.

We consider the quasi-classical Green functions 
$g(\omega_n,\theta,{\bf r})$, $f(\omega_n,\theta,{\bf r})$ and 
$f^\dagger(\omega_n,\theta,{\bf r})$ with the Matsubara frequency 
$\omega_n=(2n+1)\pi T$, where ${\bf r}$ is the center of mass 
coordinate of a Cooper pair. 
The direction of the relative momentum of the Cooper pair, 
$\hat{\bf k}={\bf k}/|{\bf k}|$, is denoted by an angle $\theta$ 
measured from the $x$ axis in the hexagonal plane. 
For the quasi-classical Green functions, the Eilenberger equation is given as  
%%%
\begin{equation}
\Bigl\{ \omega_n +{1 \over 2}
\Bigl(\partial_\parallel 
+ i\frac{2\pi}{\phi_0} A_\parallel \Bigr) \Bigr\} f
(\omega_n,\theta,{\bf r})
= \Delta({\bf r}) g(\omega_n,\theta,{\bf r}) ,
\label{eq:2.3}
\end{equation}
%%%
\begin{equation}
\Bigl\{ \omega_n -{1 \over 2}
\Bigl(\partial_\parallel 
-i\frac{2\pi}{\phi_0} A_\parallel \Bigr) \Bigr\}
f^\dagger(\omega_n,\theta,{\bf r}) 
= \Delta^\ast({\bf r}) g(\omega_n,\theta,{\bf r}),
\label{eq:2.4}
\end{equation}
%%%
\begin{equation}
\partial_\parallel g(\omega_n,\theta,{\bf r})
=\Delta^\ast({\bf r}) f(\omega_n,\theta,{\bf r}) 
-\Delta({\bf r}) f^\dagger(\omega_n,\theta,{\bf r}), 
\label{eq:2.5}
\end{equation}
%%%
\begin{equation}
g(\omega_n,\theta,{\bf r})
=\Bigl(1- f(\omega_n,\theta,{\bf r})
    f^\dagger(\omega_n,\theta,{\bf r}) \Bigr)^{1/2}, \quad 
{\rm Re} g(\omega_n,\theta,{\bf r}) > 0 ,
\label{eq:2.6}
\end{equation}
%%%
where $ \partial_\parallel = d/dr_\parallel$ and 
$A_\parallel=\hat{\bf k}\cdot {\bf A} 
=-\frac{1}{2}H r_\perp + \hat{\bf k}\cdot {\bf a}$.
Here, we have taken the coordinate system: 
$\hat{\bf u}=\cos\theta \hat{\bf x}+\sin\theta \hat{\bf y}$, 
$\hat{\bf v}=-\sin\theta \hat{\bf x}+\cos\theta \hat{\bf y}$, 
thus a point ${\bf r}=x \hat{\bf x}+y \hat{\bf y}$ 
is denoted as 
${\bf r}=r_\parallel \hat{\bf u}+r_\perp \hat{\bf v}$.
The first-order differential equations (\ref{eq:2.3})-(\ref{eq:2.5}) 
are solved along the trajectory where $r_\perp$ is held constant.

    The self-consistent conditions for the pair potential $\Delta({\bf r})$ 
and the vector potential are, respectively, given as 
%%%
\begin{equation}
\Delta({\bf r})=V N_0 
2 \pi T \sum_{\omega_n>0} \int_0^{2\pi}{d\theta \over 2\pi}
f(\omega_n,\theta,{\bf r}) ,
\label{eq:2.7}
\end{equation}
%%% 
%%%
\begin{equation}
 \nabla\times\nabla\times{\bf A}({\bf r})
= \nabla\times\nabla\times{\bf a}({\bf r})
= - \frac{\pi}{\kappa^2} 2\pi T \sum_{\omega_n>0} 
\int_0^{2\pi}{d\theta \over 2\pi}{\hat{\bf k} \over i}
g(\omega_n,\theta,{\bf r}) , 
\label{eq:2.8}
\end{equation}
%%%
where $N_0$ is the density of states at the Fermi surface, 
$V$ the pairing interaction, and 
$\kappa=(7 \zeta(3)/72)^{1/2}(\Delta_0/k_B T_c)\kappa_0=0.603\kappa_0$  
with Rieman's zeta function $\zeta(3)$. 
The GL parameter $\kappa_0$ is given by 
%%%
\begin{equation}
\kappa_0^{-2}=\frac{7 \pi \zeta(3)}{9}N_0 \left(\frac{e}{\hbar c}\right)^2 
\frac{(\hbar v_F)^4}{(\pi k_B T_c)^2} .
\label{eq:2.9}
\end{equation}
%%%
In our calculation, we use the relation 
%%%
\begin{equation}
\frac{1}{V N_0}=\ln \frac{T}{T_c}
+2\pi T \sum_{0<\omega_n < \omega_c} \frac{1}{|\omega_n|}, 
\label{eq:2.10}
\end{equation}
%%%
and set the energy cutoff $\omega_c=20 T_c$.
When ${\bf A}({\bf r})$ is calculated from the differential equation 
(\ref{eq:2.8}), the Fourier transformation is employed by following 
the method of Klein.\cite{Klein87}

We calculate the r.h.s. of Eqs. (\ref{eq:2.7}) and (\ref{eq:2.8}) 
using the quasi-classical Green functions obtained by Eqs. 
(\ref{eq:2.3})-(\ref{eq:2.6}), and obtain the 
new value for $\Delta({\bf r})$ and ${\bf A}({\bf r})$. 
Using the renewed pair potential and vector potential, 
we solve the Eilenberger equation (\ref{eq:2.3})-(\ref{eq:2.6}) again. 
Starting from the initial form, 
%%%
\begin{equation}
\Delta({\bf r})= \left(\frac{2 a_y}{a_x}\right)^{1/4}\sum_{p=-\infty}^\infty 
\exp\left\{ -\pi \frac{a_y}{a_x} \left(\frac{y+y_0}{a_y}+p \right)^2 
-2 \pi i \left[ p \left( \frac{x_0}{a_x} + \frac{\zeta}{2}p \right) 
+ \left( \frac{y_0}{a_y} + p \right) \right] \right\} 
\exp \left( -i\pi \frac{xy}{a_x a_y} \right) 
\label{eq:2.11}
\end{equation}
%%%
and ${\bf a}({\bf r})=0$, we repeat this simple iteration procedure more 
than 20 times, and obtain a sufficiently self-consistent solution for 
$\Delta({\bf r})$ and ${\bf A}({\bf r})$. 
In Eq. (\ref{eq:2.11}), the r.h.s. is the Abrikosov solution of the 
vortex lattice, where we use the relation $Ha_x a_y / \phi_0 =1$. 
The factor $\exp(-i\pi xy /a_x a_y)$ is due to the gauge 
transformation from the Landau gauge to the symmetric gauge. 
We set ${\bf r}_0=(x_0,y_0)= -\frac{1}{2}({\bf r}_1 + {\bf r}_2) $ 
so that one of the vortex centers locates at the origin of the coordinate.

On the other hand, the LDOS for energy $E$ is given by
%%%
\begin{equation}
 N(E,{\bf r})=\int_0^{2\pi}{d\theta \over 2\pi}N(E,{\bf r},\theta)
=\int_0^{2\pi}{d\theta \over 2\pi}
{\rm Re}\ g(i\omega_n \rightarrow E+i\eta,\theta,{\bf r}) , 
\label{eq:2.12}
\end{equation}
%%%
where $\eta$ is a positive infinitesimal constant, 
$N(E,{\bf r},\theta)$ the angular-resolved LDOS. 
Typically, we choose $\eta=0.01$. 
To obtain $ g(i\omega_n \rightarrow E+i\eta,\theta,{\bf r})$, 
we solve Eqs.(\ref{eq:2.3})-(\ref{eq:2.6}) for $\eta-iE$ instead 
of $\omega_n$ using the self-consistently obtained pair potential 
and vector potential.

In our calculations, we discretize ${\bf r}=u{\bf r}_1+v{\bf r}_2$ 
in a unit cell ($-0.5 \le u \le 0.5$, $-0.5 \le v \le 0.5$) for 
$\Delta({\bf r})$ and ${\bf a}({\bf r})$. 
There, $u$ and $v$ are discretized into the 41 mesh points, respectively. 
When we solve Eqs.(\ref{eq:2.3})-(\ref{eq:2.6}), we need to know 
$\Delta({\bf r})$ and ${\bf a}({\bf r})$ for arbitrary ${\bf r}$. 
It is given by interpolation of the value on the mesh point. 
It is enough that $\Delta({\bf r})$ and ${\bf a}({\bf r})$ are 
calculated within a unit cell. 
Once we obtain $\Delta({\bf r})$ and ${\bf a}({\bf r})$ in a unit cell, 
we can know $\Delta({\bf r})$ and ${\bf a}({\bf r})$ 
in other unit cells by the following relation of the 
lattice translation ${\bf R}=m{\bf r}_1+n{\bf r}_2$ ($m,n$: integer), 
respectively,  
%%%
\begin{equation}
\Delta({\bf r}+{\bf R})=\Delta({\bf r}) e^{i \chi({\bf r},{\bf R})}, \qquad 
{\bf a}({\bf r}+{\bf R})={\bf a}({\bf r}), 
\label{eq:2.13}
\end{equation}
%%%
where 
%%%
\begin{equation}
\chi({\bf r},{\bf R})=-\frac{2 \pi}{\phi_0}{\bf A}({\bf R})\cdot{\bf r}
-\pi mn +\frac{2 \pi}{\phi_0}({\bf H}\times{\bf r}_0)\cdot{\bf R} 
\label{eq:2.14}
\end{equation}
%%%
in the symmetric gauge.

Since we need a lot of CPU time for the integrating process of Eqs. 
(\ref{eq:2.3})-(\ref{eq:2.5}) in the calculation of the quasi-classical 
Green functions, we want to shorten this process as much as possible. 
It becomes possible by the symmetry consideration, as suggested by 
Klein.\cite{Klein87} 
Without the integrating process of all $\theta$ and ${\bf r}$ cases,  
we can obtain the Green functions for all $\theta$ and ${\bf r}$.  

When one of the vortex center locates at the origin of the coordinate 
(${\bf r}_0 = -\frac{1}{2}({\bf r}_1+{\bf r}_2)$), the pair potential 
has the relation 
%%%
\begin{equation}
\Delta({\bf r})=-\Delta(-{\bf r}). 
\label{eqa:2.15} 
\end{equation}
%%%
Considering the transformation ${\bf r} \rightarrow -{\bf r}$, we obtain 
the following relation from Eqs. (\ref{eq:2.3})-(\ref{eq:2.6}) and 
(\ref{eqa:2.15}), 
%%%
\begin{eqnarray}
&& f(\omega_n,\theta,{\bf r})
= -f^{\dagger \ast} (\omega_n^\ast,\theta,-{\bf r}), 
\nonumber \\ 
&& f^\dagger (\omega_n,\theta,{\bf r})
= -f^\ast (\omega_n^\ast,\theta,-{\bf r}),  
\nonumber \\
&& g(\omega_n,\theta,{\bf r})=g^\ast (\omega_n^\ast,\theta,-{\bf r}). 
\label{eqa:2.16} 
\end{eqnarray}
%%%
In the calculation of the Matsubara frequency $\omega_n$ or the case $E=0$, 
once we calculate the Green functions for ${\bf r}$ in a half area of 
a unit cell from Eqs. (\ref{eq:2.3})-(\ref{eq:2.6}), we can know the 
Green functions for all ${\bf r}$ in the unit cell from the 
relation (\ref{eqa:2.16}).

When we consider the reflection at the $x$ axis $S{\bf r}=(x,-y)$,  
our definition of $\Delta({\bf r})$ gives the relation 
%%%
\begin{equation}
\Delta({\bf r})=-\Delta^\ast(S{\bf r}). 
\label{eqa:2.17}
\end{equation}
%%%
Then we obtain the relation 
%%%
\begin{eqnarray}
&& f(\omega_n,\theta,{\bf r})
= -f^\ast (\omega_n,-\theta,S{\bf r}),
\nonumber \\
&& f^\dagger (\omega_n,\theta,{\bf r})
= -f^{\dagger \ast} (\omega_n,-\theta,S{\bf r}),  
\nonumber \\
&& g(\omega_n,\theta,{\bf r})=g^\ast (\omega_n,-\theta,S{\bf r}) 
\label{eqa:2.18}
\end{eqnarray}
%%%
from Eqs. (\ref{eq:2.3})-(\ref{eq:2.6}). 
On the other hand, there is a relation 
%%%
\begin{equation}
\Delta({\bf r})=\Delta(R^n{\bf r})e^{in \pi /3}, 
\label{eqa:2.19}
\end{equation}
%%%
when we consider the rotation $n \pi /3$ ($n$: integer) around the origin 
of the coordinate; $R^n{\bf r}=(x \cos(n \pi /3) -y \sin(n \pi /3), 
x \sin(n \pi /3) + y \cos(n \pi /3) )$. 
Then we obtain the relation
%%%
\begin{eqnarray}
&& f(\omega_n,\theta,{\bf r})
= f(\omega_n,\theta + n \pi/3,R^n{\bf r})e^{i n \pi /3},
\nonumber \\
&& f^\dagger (\omega_n,\theta,{\bf r})
= f^\dagger(\omega_n,\theta+ n \pi/3,R^n{\bf r})e^{-i n \pi /3},  
\nonumber \\
&& g(\omega_n,\theta,{\bf r})=g(\omega_n,\theta+ n \pi/3,R^n{\bf r}) 
\label{eqa:2.20}
\end{eqnarray}
%%%
from Eqs. (\ref{eq:2.3})-(\ref{eq:2.6}). 
Once we calculate the Green functions for $0 \le \theta \le \pi /6$ from 
Eqs. (\ref{eq:2.3})-(\ref{eq:2.6}), we can know the Green functions for all 
$\theta$ from the relations (\ref{eqa:2.18}) and (\ref{eqa:2.20}).

There are two methods to solve the Eilenberger equation 
(\ref{eq:2.3})-(\ref{eq:2.6}); the explosion method and the symmetry method. 
We describe them in the following. 

%%%% s2.1 %%%%%%%%%%%%%
\subsection{explosion method}

We use the so-called explosion method\cite{Klein87,Thuneberg} in this 
paper to obtain the quasi-classical Green function  
%%%
\begin{equation}
\hat g \equiv 
\left( \begin{array}{cc} g & i  f \\ 
-i f^\dagger &-g \end{array}\right) 
\label{eqa:2.21}
\end{equation}
%%%
from the Eilenberger equation (\ref{eq:2.3})-(\ref{eq:2.6}).  
In addition to a physical solution $\hat g_{\rm ph}$,
Eqs. (\ref{eq:2.3})-(\ref{eq:2.6}) have two unphysical solutions 
$\hat g_+ $ and $\hat g_-$. 
The solutions $\hat g_\pm$ explode (increase exponentially) 
in the directions $\pm {\bf k}$ and decrease in the opposite directions. 
Even when we use the physical solution as an initial value, 
the unphysical solutions always mix and become dominant during 
the process of the numerical integration of 
Eqs. (\ref{eq:2.3})-(\ref{eq:2.5}) along a long path. 
We obtain $\hat g_\pm$ by integrating 
from $r_\parallel \mp r_{\rm A}$ to $ r_\parallel $, 
where $r_{\rm A}(>0)$ is large so that explosion takes place. 
It is known\cite{Klein87,Thuneberg} that the physical solution 
is obtained from the commutator of the two unphysical solutions,  
%%%
\begin{equation} 
\hat g_{\rm ph}= c [ \hat g_+ , \hat g_- ], 
\label{eqa:2.22}
\end{equation}
%%%
where $c$ is a constant determined from Eq. (\ref{eq:2.6}). 
In this method, we can obtain the solution for arbitrary 
${\bf k}$ directions. 

%%%% s2.2 %%%%%%%%%%%%%
\subsection{symmetry method}

In the case of solving the Eilenberger equation in the real energy $E$, 
Klein could not succeed in the explosion method. 
Then, he calculated the angular-resolved LDOS by means of the 
so-called symmetry method.\cite{Klein89,Pottinger}
The lattice translation ${\bf R}=m{\bf r}_1 +n{\bf r}_2$ 
($m,n$: integer) of the quasi-classical Green function is given as 
%%%
\begin{eqnarray}
&& f(\omega_n,\theta,{\bf r}+{\bf R})
=f(\omega_n,\theta,{\bf r})e^{i \chi({\bf r},{\bf R})},
\nonumber \\
&& f^\dagger(\omega_n,\theta,{\bf r}+{\bf R})
=f^\dagger(\omega_n,\theta,{\bf r})e^{-i \chi({\bf r},{\bf R})}, 
\nonumber \\
&& g(\omega_n,\theta,{\bf r}+{\bf R})=g(\omega_n,\theta,{\bf r}), 
\label{eqa:2.23}
\end{eqnarray}
%%%
where $\chi({\bf r},{\bf R})$ is defined in Eq. (\ref{eq:2.14}). 
When we solve Eqs. (\ref{eq:2.3})-(\ref{eq:2.6}) for the ${\bf k}$ 
direction parallel to $m{\bf r}_1 +n{\bf r}_2$, the solution have to 
satisfy Eq. (\ref{eqa:2.23}) as the boundary condition. 
Thus, we integrate the first differential equations 
(\ref{eq:2.3})-(\ref{eq:2.5}) along the trajectory 
from ${\bf r}$ to ${\bf r}+{\bf R}$, and search the solution which 
satisfies the boundary condition (\ref{eqa:2.23}) 
by the so-called shooting method or other method. 

Compared with the explosion method, the integral path can be short 
in the symmetry method, especially for the calculation of the real 
energy $E$.  
Then the CPU time of the numerical calculation may be short. 
However, in the symmetry method, we can obtain the solution only for 
the specific ${\bf k}$ direction parallel to $m{\bf r}_1 +n{\bf r}_2$. 
Since we must know the solution for all ${\bf k}$ directions to 
calculate the LDOS in Eq. (\ref{eq:2.12}), we cannot obtain 
the LDOS in this method.

%%%% s3 %%%%%%%%%%%%%%%%%%%%%%%%%%%%%%%%%%%%%%%%%%%%%%%%%%%%%%%
\section{Pair potential, magnetic field and LDOS}
\label{sec:3}

The pair potential, the vector potential and the LDOS are calculated for the 
material parameter appropriate to ${\rm NbSe_2}$, i.e., the BCS coherence 
length 77 ${\rm \AA}$ and the BCS penetration depth 690 
${\rm \AA}$.\cite{Hess3} 
As an example of a low magnetic field, we consider the case  
$\bar{H}$= 0.1 Tesla ($a_x=6.4 \xi$).  
And as an example of a high magnetic field, we consider the case
$\bar{H}$= 1 Tesla ($a_x=2.0 \xi$). 
As for the example of further high field, the cases 
$\bar{H}$= 2 Tesla ($a_x=1.4 \xi$) and 4 Tesla ($a_x=1.0 \xi$) 
are also considered. 
Figure \ref{fig:1} presents the configuration of the vortex lattice 
presented in figures in this paper. 
In the figures, there are 7 vortex centers, and one of the vortex centers 
locates at the center of each figure. 
When we consider the profile of the spatial variation, we present it
along the trajectories of line OA ($0^\circ$ radial direction),
line OB ($15^\circ$ radial direction), line OC ($30^\circ$ radial direction)
and line AC (boundary).

%%%% s3.1 %%%%%%%%%%%%%
\subsection{Pair potential and magnetic field}

The amplitude of the self-consistently obtained pair potential 
$|\Delta({\bf r})|$ is shown in the contour plot of Fig. \ref{fig:2}. 
For $\bar{H}$=0.1 Tesla in Fig. \ref{fig:2} (a), the core region 
localizes in the small region around the vortex center in the unit cell. 
For $\bar{H}$=1 Tesla in Fig. \ref{fig:2} (b), the core region occupies 
the large part of the unit cell. 
Then the core regions substantially overlap each other. 
As seen from the contour lines, 
while the inner region of the vortex core (the region where 
$|\Delta({\bf r})|<0.8 $) has cylindrically symmetric structure, 
the outer region ($|\Delta({\bf r})| \ge 0.8$) shows the sixfold 
symmetric structure. 
There, the amplitude along the nearest-neighbor vortex direction 
($0^\circ$ direction) is a little suppressed compared with that along the 
next nearest-neighbor vortex direction ($30^\circ$ direction).
In this high field case, the maximum of the amplitude is suppressed 
down to 0.95$\Delta_0$. 
The sixfold symmetric structure of $|\Delta({\bf r})|$ appears more clearly 
in the case $\bar{H}$=2 Tesla in Fig. \ref{fig:2} (c), where the maximum 
of the amplitude is suppressed down to 0.83$\Delta_0$. 
In the case $\bar{H}$=4 Tesla, the pair potential  has similar sixfold 
symmetric spatial distribution, but its amplitude is suppressed down 
to 0.38$\Delta_0$.

The spatial variation of the magnetic field is obtained from the 
self-consistently obtained vector potential. 
The microscopic magnetic field $H({\bf r})$ is shown in Fig. \ref{fig:3}. 
For $\bar{H}$=0.1 Tesla in Fig. \ref{fig:3} (a), $H({\bf r})$ has 
a cylindrically symmetric sharp peak at each vortex center. 
The variation of $H({\bf r})$ is in the range 
$0.75 \le H({\bf r})/\bar H \le 2.95$. 
For $\bar{H}$=1 Tesla in Fig. \ref{fig:3} (b), this peak becomes lower 
and broader ($0.981 \le H({\bf r})/\bar H \le 1.096$). 
It has almost cylindrically symmetric but slightly sixfold symmetric 
structure. 
There $H({\bf r})$ extends slightly to the $30^\circ$ direction in the 
core region.
The sixfold symmetric distribution of $H({\bf r})$ is seen clearly 
in the case $\bar{H}$=2 Tesla in Fig. \ref{fig:3} (c), where 
the peak of $H({\bf r})$ becomes further low and broad structure 
($0.994 \le H({\bf r})/\bar H \le 1.025$).

%%%% s3.2 %%%%%%%%%%%%%
\subsection{Sixfold symmetric LDOS}

The LDOS in Eq. (\ref{eq:2.12}) is calculated by using the self-consistently 
obtained pair potential and vector potential presented in Figs. 
\ref{fig:2} and \ref{fig:3}. 

As for the angular-resolved LDOS $N(E,{\bf r},\theta)$ in Eq. 
(\ref{eq:2.12}), the results of Klein in the symmetry method\cite{Klein89} 
are qualitatively reproduced by our calculation of the explosion 
method. 
In the single vortex case, $N(E,{\bf r},\theta)$ consists of a 
straight sharp ridge line parallel to the ${\bf k}$-direction, where its 
distance from the vortex center increases on raising $E$. 
At a high magnetic field of the vortex lattice case, the ridge of  
$N(E,{\bf r},\theta)$ extends to the neighboring unit cells (see, for example, 
Fig. 5 in Ref. \onlinecite{Klein89}). 
Especially it is noted that, 
when ${\bf k}$ points toward the nearest-neighbor vortex direction ($\theta
=0^\circ$ or $\hat {\bf k}_{10}$ in the notation of Klein\cite{Klein89}), 
the sharp ridge in the single vortex case splits into two parallel ridges.  
There, $N(E,{\bf r},\theta)$ distributes broadly in the region between 
them (see Fig. 10 in Ref. \onlinecite{Klein89}). 
Also when ${\bf k}$ points toward the next nearest-neighbor vortex direction 
($\theta=30^\circ$ or $\hat {\bf k}_{11}$), $N(E,{\bf r},\theta)$ slightly 
splits into two ridges. 
The above-mentioned splitting of the ridge occurs more clearly 
for higher magnetic field or larger $E$. 
At a weak magnetic field, the ridge of $N(E,{\bf r},\theta)$ localizes 
within a unit cell, and the splitting of the ridge does not occur even 
for $\theta=0$ (see Fig. 1 in Ref. \onlinecite{Klein89}).

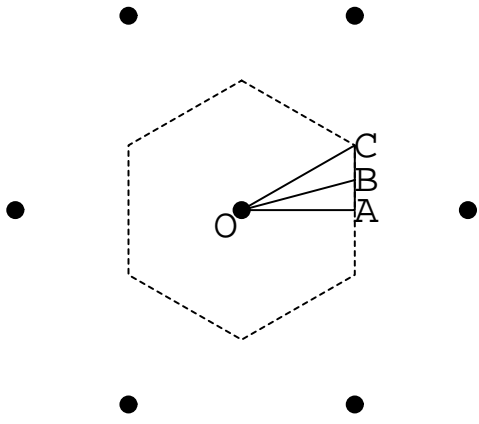
\begin{figure}
\caption{
The configuration of the vortex lattice in our figures. 
The vortex centers are shown by $\bullet$. 
A hexagon enclosed by dashed lines presents the Wigner-Seitz cell of 
the vortex lattice. 
When we consider the profile of the spatial variation, we present it 
along the trajectories of line OA ($0^\circ$ radial direction), 
line OB ($15^\circ$ radial direction), line OC ($30^\circ$ radial direction) 
and line AC (boundary).  
}
\label{fig:1}
\end{figure}

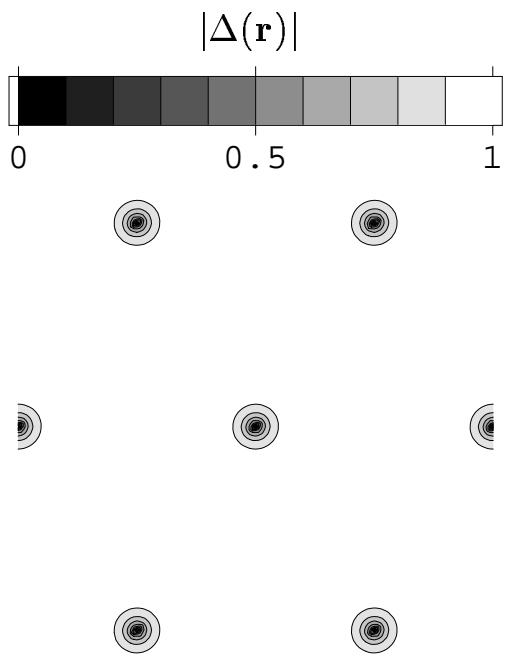
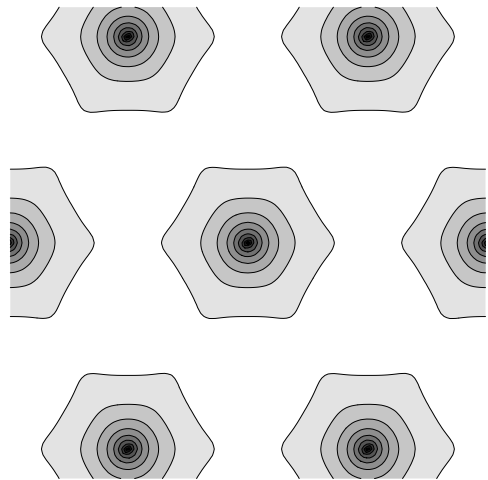
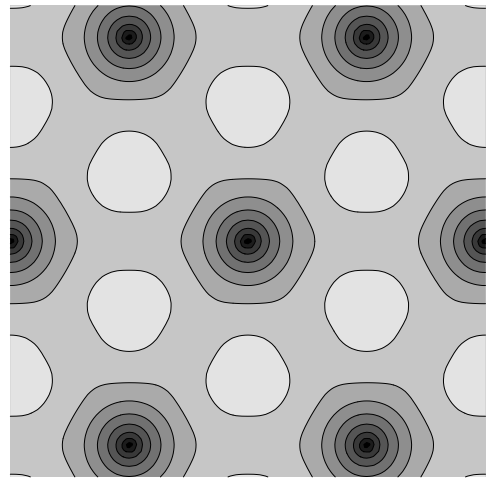
\begin{figure}
\caption{
Spatial variation of the pair potential at $T/T_c= 0.1$. 
Contour plot of the amplitude, $|\Delta({\bf r})|$, is presented. 
(a) $\bar H= 0.1$ Tesla. The region $12.8 \xi \times 12.8 \xi $ is presented. 
The core region localizes in a small area around each vortex center. 
(b) $\bar H=1$ Tesla. The region $4 \xi \times 4 \xi $ is presented.
(c) $\bar H=2$ Tesla. The region $2.8 \xi \times 2.8 \xi $ is presented.
For $\bar H=$1 Tesla and 2 Tesla, the core region occupies the large part 
of the unit cell. 
The pair potential shows cylindrical symmetry at the inner region of 
the vortex core, and sixfold symmetry at the outer region. 
}
\label{fig:2}
\end{figure}

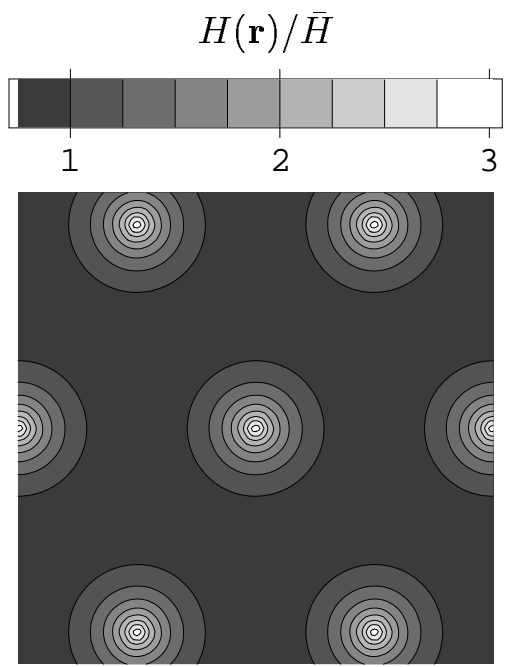
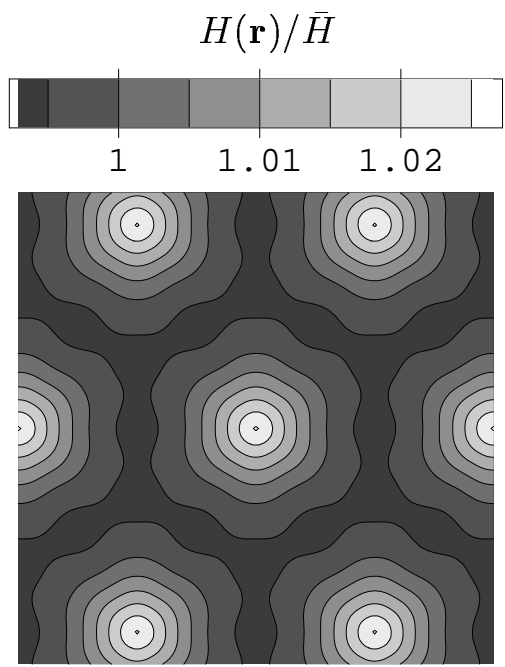
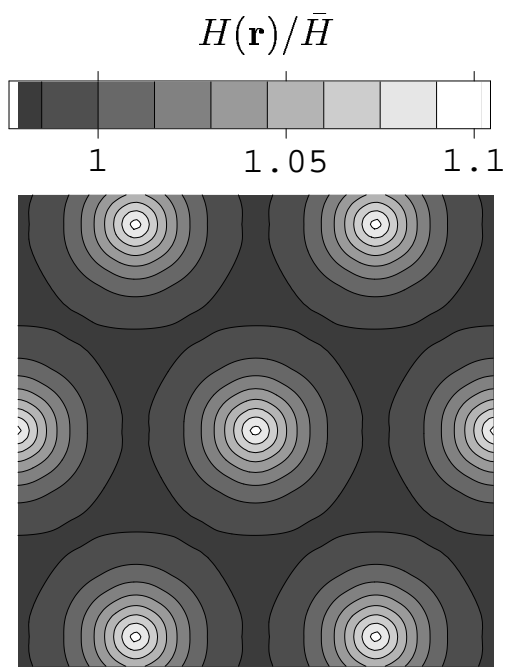
\begin{figure}
\caption{
Spatial variation of the magnetic field at $T/T_c= 0.1$. 
Contour plot of the normalized value $H({\bf r})/\bar H $ is presented 
for $\bar H = 0.1$ Tesla (a), 1 Tesla (b) and 2 Tesla (c). 
For $\bar H=1$ Tesla and 2 Tesla, $H({\bf r})$ extends a little 
to the $30^\circ$ direction in the vortex core region. 
}
\label{fig:3}
\end{figure}

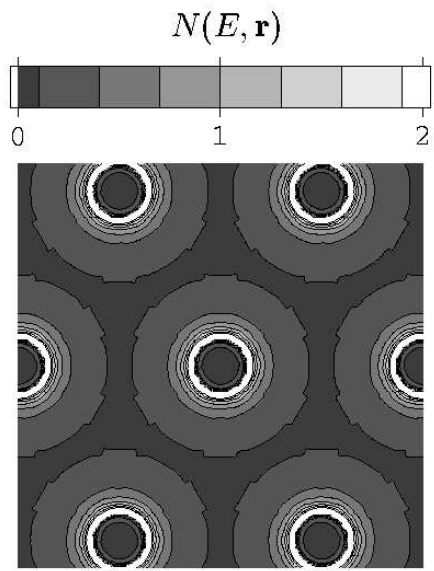
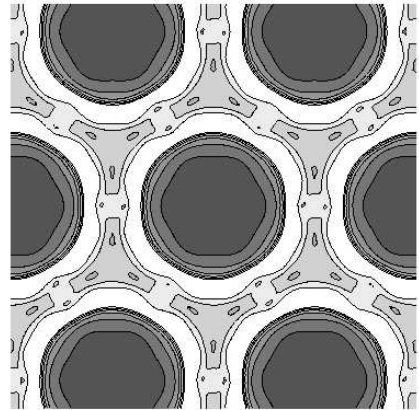
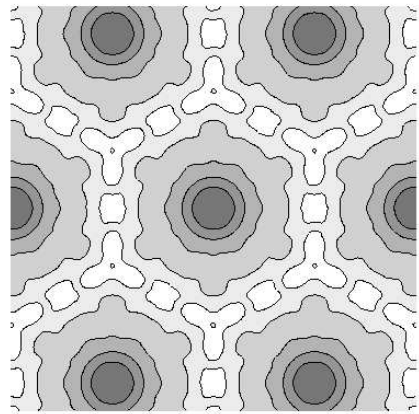
\begin{figure}
\caption{
Spatial variation of the LDOS $N(E, {\bf r})$ at $\bar H= 0.1$ Tesla. 
The contour plot is shown for $E$=0.8 (a), 1.0 (b) and 1.2 (c). 
The LDOS structure shows cylindrical symmetry for lower energy $E \le 0.8$, 
and sixfold symmetry for higher energy $E \ge 1$. 
}
\label{fig:4}
\end{figure}

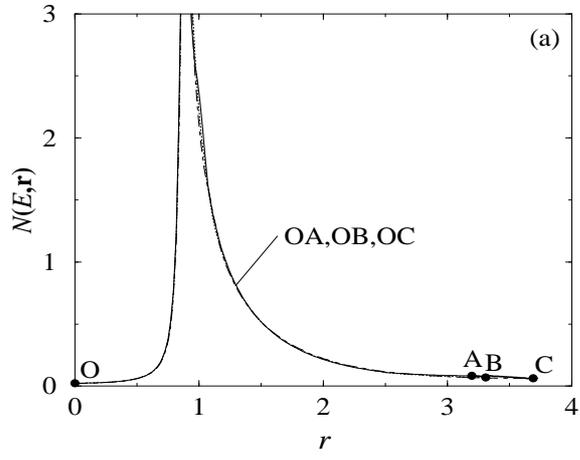
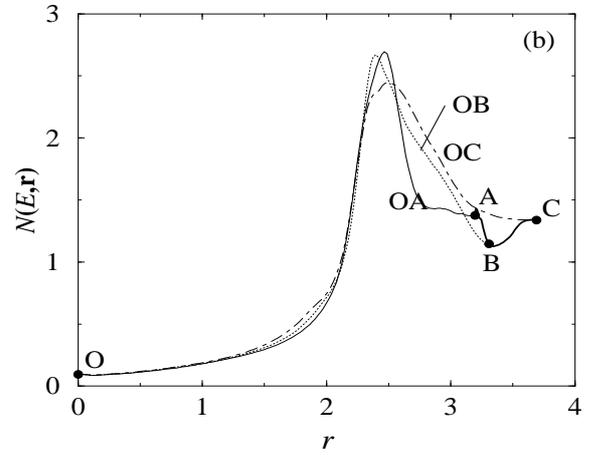
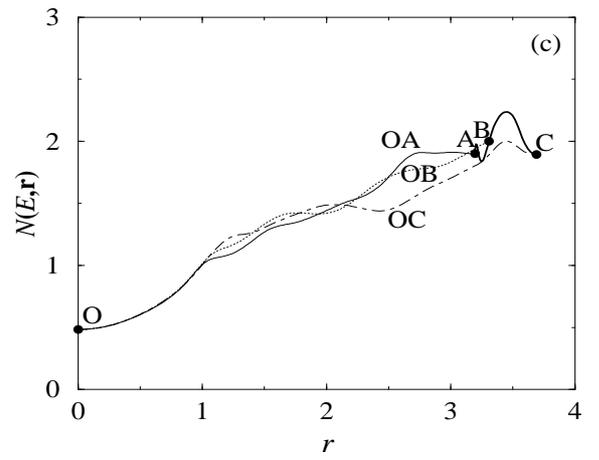
\begin{figure}
\caption{
Spatial variation of the LDOS $N(E, {\bf r})$ at $\bar H= 0.1$ Tesla.
The profiles along the lines OA (solid line), OB (dotted line), OC 
(dot-dashed line) and AC (thick line) of Fig. 1 are presented for 
$E$=0.8 (a), 1.0 (b) and 1.2 (c) 
as a function of $r$ (distance from the vortex center). 
Each figure corresponds to that of Fig. 4.
}
\label{fig:5}
\end{figure}

Integrating $N(E,{\bf r},\theta)$ over all $\theta$, we obtain the LDOS 
$N(E,{\bf r})$ in Eq. (\ref{eq:2.12}). 
The spatial variation of the numerically obtained LDOS 
is shown in Figs. \ref{fig:4} and \ref{fig:5}  for the weak 
magnetic field case $\bar{H}=0.1$ Tesla. 
Figure \ref{fig:4} shows the contour plot of $N(E,{\bf r})$ for each $E$. 
Figure \ref{fig:5} shows the profiles of $N(E,{\bf r})$ along the lines 
OA, OB, OC and AC of Fig. \ref{fig:1}. 
As shown in Fig. \ref{fig:4} (a),  the LDOS shows cylindrical structure 
in the low energy case ($E \le 0.8$), where the ridge of the LDOS forms 
a ring around each vortex center.  
It is the similar structure as that obtained in the single vortex case 
for $s$-wave pairing (see, for example, Fig. 1 in Ref. \onlinecite{Schopohl}). 
However, in the higher energy case shown in Figs. \ref{fig:4} (b) 
and (c), the radius of the ring approaches the boundary of a unit cell 
in the vortex lattice, and the LDOS becomes sixfold symmetric structure.  

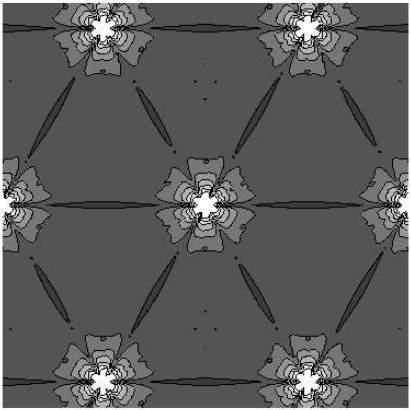
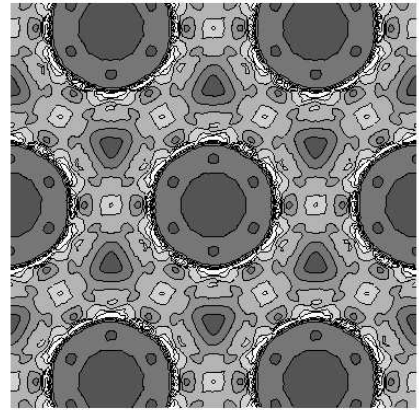
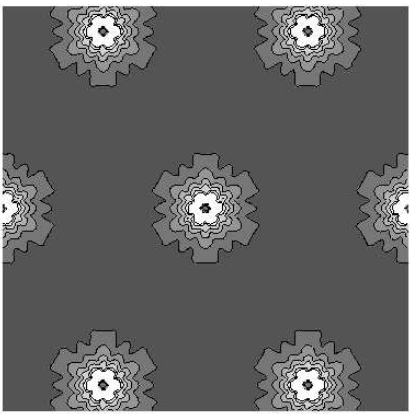
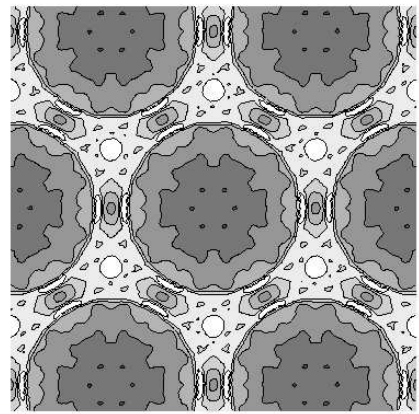
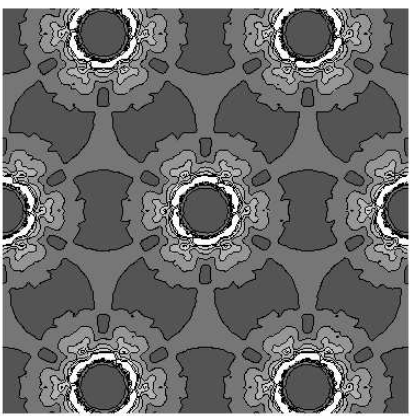
\begin{figure}
\caption{
Spatial variation of the LDOS $N(E, {\bf r})$ at $\bar H= 1$ Tesla. 
The contour plot is shown for $E$=0 (a), 0.2 (b), 0.6 (c), 1.0 (d) 
and 1.2 (e). 
In this high magnetic field case, the LDOS shows sixfold symmetry even at 
the low energy down to $E=0$. 
}
\label{fig:6}
\end{figure}

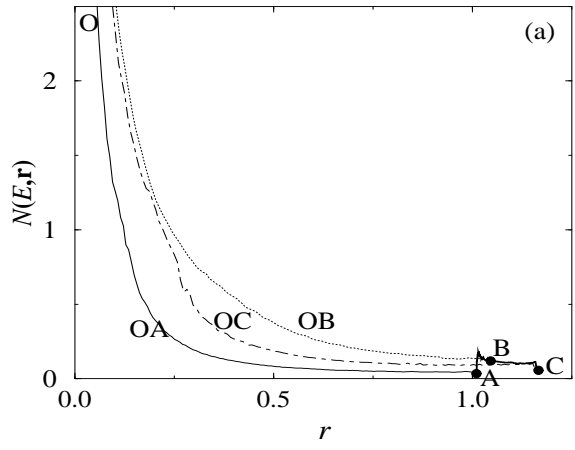
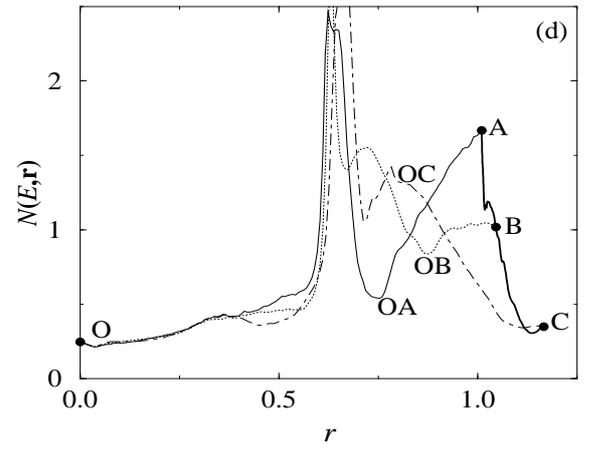
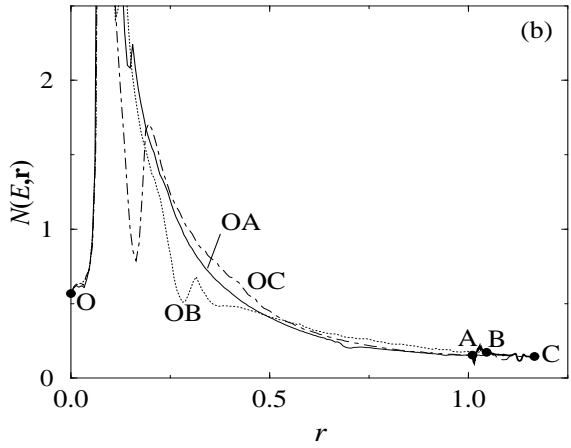
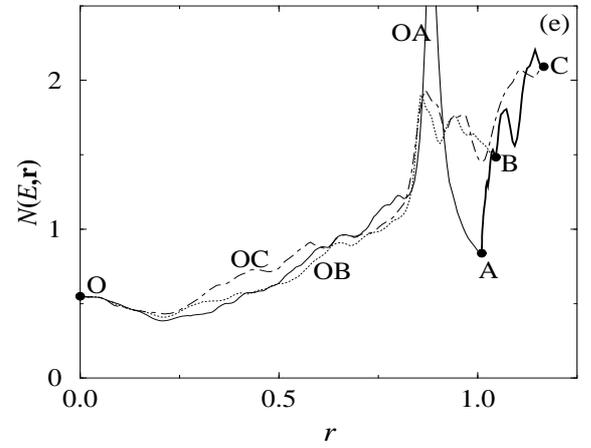
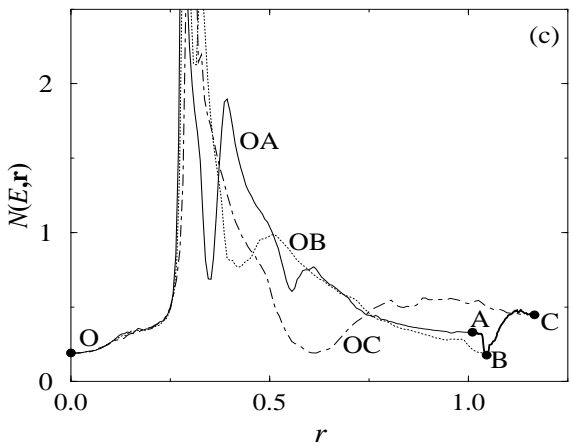
\begin{figure}
\caption{
Spatial variation of the LDOS $N(E, {\bf r})$ at $\bar H= 1$ Tesla. 
The profiles along the lines OA (solid line), OB (dotted line), OC 
(dot-dashed line) and AC (thick line) of Fig. 1 are  
presented for $E$=0 (a), 0.2 (b), 0.6 (c), 1.0 (d) and 1.2 (e) 
as a function of $r$.  
Each figure corresponds to that of Fig. 6.
}
\label{fig:7}
\end{figure}

The LDOS for the high field case $\bar{H}=1$ Tesla is shown in Figs. \ref{fig:6} and \ref{fig:7}. 
It presents the sixfold symmetric star-shaped structure 
even in the low energy case. 
At $E=0$, the ray of the star extends to the next nearest-neighbor 
directions (the $30^\circ$ directions) as presented in Fig. \ref{fig:6} (a). 
The width of the ray is rather broad compared with the experimental 
data.\cite{Hess2,Hess3,Hess4,Hess5} 
At higher energy $E \sim 0.6$, on the other hand, the ray extends to the 
nearest-neighbor directions (the $0^\circ$ directions) as shown in 
Fig.  \ref{fig:6} (c). 
The orientation of the star relative to the vortex lattice is consistent 
with the experimental data by Hess {\it et al}.\cite{Hess2,Hess3,Hess4,Hess5} 
Therefore, the experimental features (1) and (2) mentioned in 
Sec. \ref{sec:1}, including the orientation of the star, 
are qualitatively reproduced without further assumptions. 
As presented in Fig.  \ref{fig:6} (b), 
at the intermediate energy of the $30^\circ$ rotation of the star, 
the amplitude of the ray  in the $30^\circ$ directions decreases 
and the new ray starts to extend in the $0^\circ$ directions 
with increasing $E$. 
This behavior is in discord with the experimental data. 
In the experiment on ${\rm NbSe_2}$, the LDOS shows the split parallel 
ray structure at the intermediate energy (the feature (3) mentioned in
Sec. \ref{sec:1}).

The experimental data around $E \sim 1$ can be qualitatively reproduced 
by our calculation. 
For $E \sim 1$, there are little LDOS distribution around the vortex 
core region, and the LDOS distributes around the boundary of the 
Wigner-Seitz cell of the vortex lattice. 
Therefore, the vortex core is detected as a dark object in the STM image.
Even in the case the vortex core localizes in the narrow region 
around the vortex center, the dark STM image of the vortex core 
for $E \sim 1$ has the size comparable to the unit cell, 
as shown in Figs. \ref{fig:4} (b) and (c). 
In the LDOS distribution around $E \sim 1$, we pay our attention to 
the value at the boundary of the Wigner-Seitz cell, that is, on the 
line AC in Fig. \ref{fig:1}. 
For $E \le 1$, $N(E,{\bf r})$ is large at the point A compared with the 
point C, as seen from Fig. \ref{fig:6} (d). 
On the other hand, the peak of $N(E,{\bf r})$ shifts to the point C 
on raising $E$ as seen from Fig. \ref{fig:6} (e), which is consistent 
with the experimental STM image for 1.2 mV in Fig. 1 of 
Ref. \onlinecite{Hess4}.  

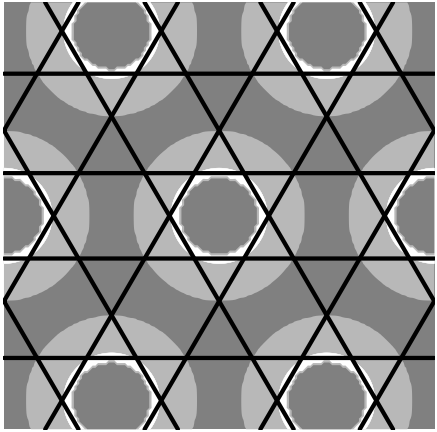
\begin{figure}
\caption{
Schematically presented LDOS structure, which corresponds to that of Fig. 6. 
Without the vortex lattice effect, the LDOS has the cylindrical structure, 
which is presented as white rings. 
To this structure, the vortex lattice effect suppresses the LDOS along 
the lines presented in the figure.
Therefore, it is seen that the LDOS in Fig. 6 has small value 
at the points along the lines. 
}
\label{fig:8}
\end{figure}

The spatial variation of the LDOS structure in Fig. \ref{fig:6} 
can be explained as follows. 
Without the vortex lattice effect, the LDOS forms the cylindrical structure 
which is shown schematically as white rings in Fig. \ref{fig:8}.  
It has the ridge on a ring around each vortex center, and the LDOS 
distributes at the outer side of the rings. 
To this distribution, the vortex lattice effect suppresses the LDOS along 
the common tangent lines on these rings (lines in Fig. \ref{fig:8}). 
These lines runs toward the $0^\circ$ and its equivalent directions  
(the nearest-neighbor vortex directions).  
This suppression is due to the splitting of the ridges in the 
angle-resolved LDOS $N(E,{\bf r},\theta)$ for $\theta$ of the 
nearest-neighbor directions. 
In the contribution to the LDOS, the ridge of $N(E,{\bf r},\theta)$ along 
the lines of Fig. \ref{fig:8} becomes broad and low-height distribution  
due to the splitting (see Fig. 10 in Ref. \onlinecite{Klein89}). 
Therefore, it is seen that the LDOS in Fig. \ref{fig:6} has small value 
at the points along the lines. 
The small suppression is also shown along the tangent lines running toward 
the $30^\circ$ and its equivalent directions (the next nearest-neighbor 
vortex directions). 
Considering the fact that the radius of the ring in Fig. \ref{fig:8} 
increases on elevating $E$, we can understand the change of the LDOS in 
Figs. \ref{fig:6} (a) to (e). 
Since the radius reduces to 0 in the limit $E \rightarrow 0$, the 
suppression lines in Fig. \ref{fig:8} reduce to the lines connecting 
nearest-neighbor vortex centers for $E=0$, which is seen in 
Fig. \ref{fig:6} (a).

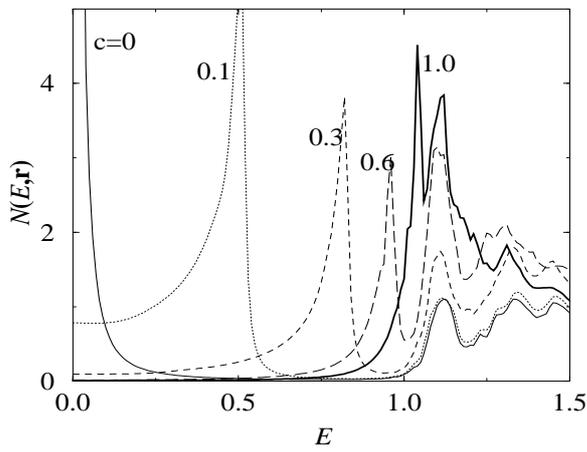
\begin{figure}
\caption{
Spectrum $N(E, {\bf r})$ at the points ${\bf r}$ along 
the $0^\circ$ radial direction line OA for $\bar H= 0.1$ Tesla.
The sample points are given by ${\bf r}=(c a_x, 0)$ with 
$c$=0, 0.1, 0.3, 0.6 and 1. 
}
\label{fig:9}
\end{figure}

Another way to examine the quasiparticle excitations in the vortex state 
is to see how the spectrum evolves along the radial directions. 
We consider the spectrum at the points along the $0^\circ$ (line OA in 
Fig. \ref{fig:1}), $15^\circ$ (line OB) and $30^\circ$ (line OC) 
radial directions. 
In the figures, we show the spectrum at ${\bf r}=(c a_x,0)$ for the 
$0^\circ$ direction, ${\bf r}=(c a_x,c a_x \tan(\pi/12))$ for the 
$15^\circ$ direction and ${\bf r}=(c a_x,c a_x \tan(\pi/6))$ for the 
$30^\circ$ direction, where we choose the point for $c$=0, 0.1, 0.3, 0.6, 1. 
As for the case $\bar{H}=0.1$ Tesla, Fig. \ref{fig:9} shows 
the spectrum at  the points along the $0^\circ$ direction. 
For $E \le 0.8$, the spectrum has the similar structure as that of the single 
vortex case (see Fig. 2 in Ref. \onlinecite{Ullah} for comparison). 
At higher energy, the peak at $E=1$ in the single vortex case becomes 
broad and shifts a little to higher energy side due to the 
vortex lattice effect. 
Along the other radial directions, the LDOS  has the almost similar 
spectrum as that of the $0^\circ$ direction. 
Small differences are shown only for $E \ge 0.8$. 

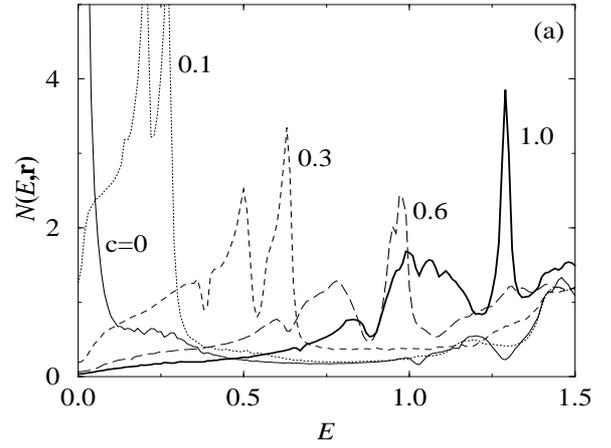
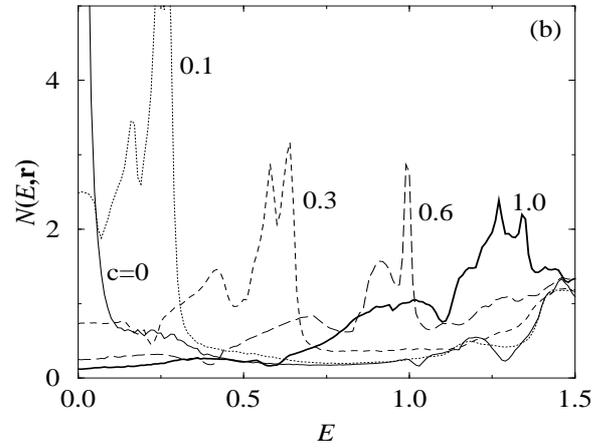
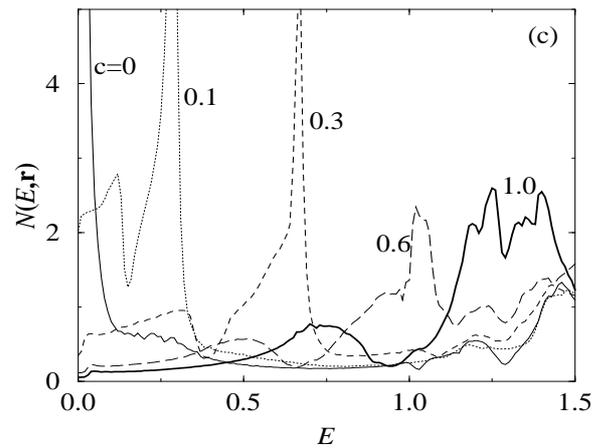
\begin{figure}
\caption{
Spectrum $N(E, {\bf r})$ at the points ${\bf r}$ along 
the $0^\circ$ radial direction line OA (a), the $15^\circ$ line OB (b) 
and the $30^\circ$ line OC (c) for $\bar H= 1$ Tesla.
The sample points are given by ${\bf r}=(c a_x, 0)$ for (a), 
$(c a_x, c a_x \tan(\pi/12))$ for (b) and 
$(c a_x, c a_x \tan(\pi/6))$ for (c), with 
$c$=0, 0.1, 0.3, 0.6 and 1. 
}
\label{fig:10}
\end{figure}

On the other hand, the spectrum for $\bar{H}=1$ Tesla is 
shown in Fig. \ref{fig:10}, where (a), (b) and (c) are that along $0^\circ$, 
$15^\circ$ and $30^\circ$ radial directions, respectively. 
It is notable that the lower energy peak in Fig. \ref{fig:9} 
splits into two or three peaks in this higher field case. 
The behavior of these peaks varies depending on the direction of the 
radial lines, as shown in Fig. \ref{fig:10} (a)-(c).
The peak above $E=1$ in Fig. \ref{fig:9} is suppressed in Fig. \ref{fig:10}.
The peak of the spectrum at the boundary point ($c=1$) shifts to 
higher energy from 1. 

%%%% s3.3 %%%%%%%%%%%%%
\subsection{LDOS at higher field}

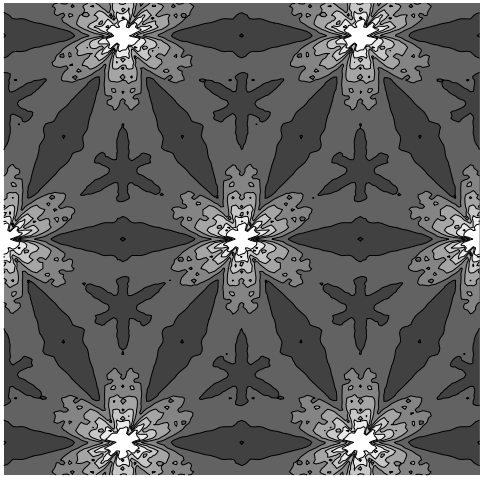
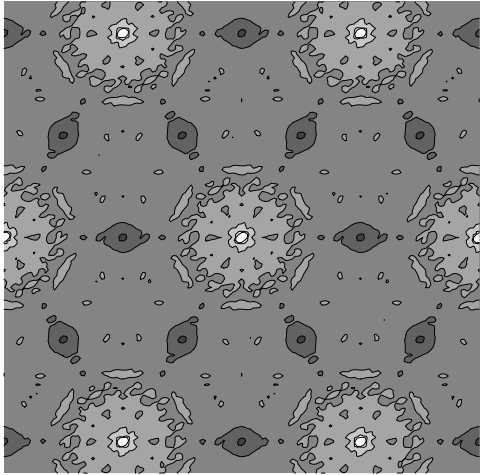
\begin{figure}
\caption{
Spatial variation of the LDOS $N(E,{\bf r})$ for $E=0$. 
Contour plot is shown for $\bar H=$2 Tesla (a) and 4 Tesla (b). 
At higher magnetic fields, the LDOS around the vortex is 
connected each other. 
}
\label{fig:11}
\end{figure}

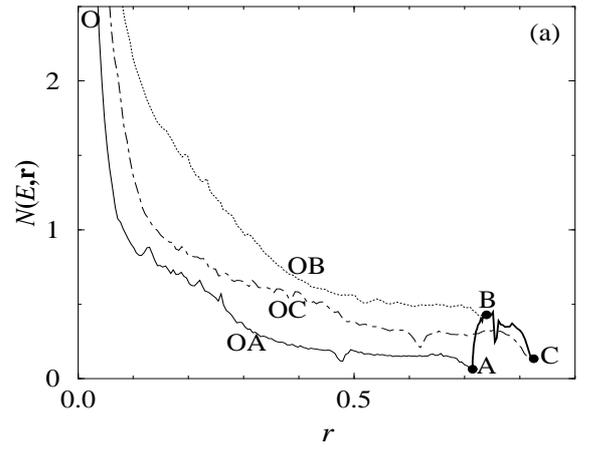
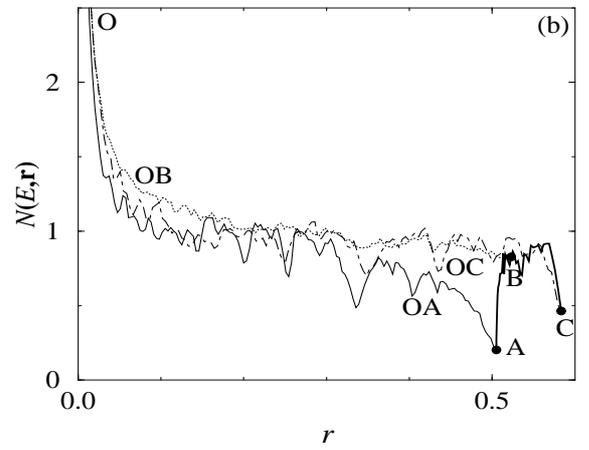
\begin{figure}
\caption{
Spatial variation of the LDOS $N(E,{\bf r})$ for $E=0$. 
Profile along the line OA (solid line), OB (dotted 
line), OC (dot-dashed line) and AC (thick line) of Fig. 1. 
are presented for $\bar H=$2 Tesla (a) and 4 Tesla (b) 
as a function of $r$. 
Each figure corresponds to that of Fig. 11.
}
\label{fig:12}
\end{figure}

Here, to discuss the transfer of the quasiparticle bound states 
between vortex cores, we focus on the LDOS structure at $E=0$. 
With increasing a magnetic field, the peak of the LDOS at the vortex center 
decreases ($N(E=0,{\bf r}=0)=$71 for 0.1 Tesla, 65 for 1 Tesla, 
44 for 2 Tesla and 11 for 4 Tesla), and the distribution of the LDOS extends 
sixfold symmetrically to the wider region. 
Therefore, the LDOS around the vortex core is connected each other at 
high magnetic field even in the zero energy state.
To show the connection clearly, the LDOS for $\bar H=2$ Tesla and 
4 Tesla is presented in Figs. \ref{fig:11} and \ref{fig:12}. 
At $\bar H=2$ Tesla in Fig. \ref{fig:11} (a), the LDOS around the 
vortex core is connected each other at the 
middle point of the line AC of Fig. \ref{fig:1}. 
At further high field $\bar H=4$ Tesla shown in Fig. \ref{fig:11} (b), 
the LDOS is seems to have the component uniformly distributing all over 
the unit cell. 
There, the LDOS has small peak at the vortex center and minimum at the 
stationary point of the current flow (point A of Fig. \ref{fig:1}). 
With raising a magnetic field, the LDOS at the boundary 
of the Wigner-Seitz cell increases (its maximum value is 
0.01 for 0.1 Tesla, 0.18 for 1 Tesla, 0.44 for 2 Tesla and 0.90 
for 4 Tesla), which means that the transfer of the quasiparticle 
bound states  between vortex cores increases. 

On ${\rm NbSe_2}$, the dHvA oscillation is observed at 
magnetic fields down to 4 Tesla in the superconducting mixed state. 
As discussed above, we show that there is large transfer of the 
bound state between vortex cores in this high field region. 
This transfer seems to be a possible origin of the dHvA oscillation 
in superconductors.

%%%% s4 %%%%%%%%%%%%%%%%%%%%%%%%%%%%%%%%%%%%%%%%%%%%%%%%%%%%%%%
\section{Summary and discussions}
\label{sec:4}

By using the self-consistently obtained pair potential and vector potential, 
the LDOS of the triangular vortex lattice is calculated in an 
isotropic $s$-wave superconductors based on the quasi-classical 
Eilenberger theory. 
Important results obtained for the vortex lattice are as follows; 
(i) We do find a sixfold star shape of the LDOS and the $30^\circ$ 
rotation upon elevating the bias energy. 
The sixfold star originates from the triangular vortex lattice effect. 
(ii) The orientation of the star coincides with the STM data, 
namely the ray extends toward the next nearest-neighbor vortex direction 
at lower bias energy. 
Therefore we succeed in determining the absolute direction relative to the 
vortex lattice. 
This is one of the most eminent features in the STM data 
by Hess {\it et al.}\cite{Hess1,Hess2,Hess3,Hess4,Hess5}  
Thus the experimental features (1) and (2) mentioned in Sec. \ref{sec:1} 
are qualitatively reproduced by the naive calculation of the LDOS in 
the triangular vortex lattice. 
(iii) For $E\sim 1$, the LDOS distributes around the boundary of the 
Wigner-Seitz cell of the vortex lattice, and there are little 
distribution around the vortex core region. 
Therefore, the vortex core is detected as a sixfold symmetric dark 
object in the STM images.
At the energy $E \sim 1.2$, the LDOS has large intensity at the point 
farthest from the vortex center (the point C in Fig. \ref{fig:1}). 
(iv) The characteristic sixfold symmetric LDOS structure in the low 
energy case appears only at a high magnetic field such as 1 Tesla 
for the material parameters appropriate to ${\rm NbSe_2}$, 
where the core regions substantially overlap each other. 
At a lower magnetic field such as 0.1 Tesla, the LDOS reduces to the 
almost cylindrical structure in the low energy case.

As mentioned above, the numerically obtained LDOS in the vortex lattice 
case can qualitatively reproduce the characteristic features (1) and 
(2) in the experimental data on ${\rm NbSe_2}$. 
However, the detailed comparison reveals some discrepancies between 
the experimental data and our results, as follows; 
(v) At the intermediate energy of the $30^\circ$ rotation, 
our results does not reproduce the split parallel ray structure 
(the feature (3) mentioned in Sec. \ref{sec:1}). 
In the experiment, the ray of the star in the $30^\circ$ direction at 
$E=0$ splits into two parallel rays with elevating $E$. 
In our results, the ray along the $30^\circ$ direction at $E=0$ weakens and 
new ray starts to extend along the $0^\circ$ direction with increasing $E$. 
(vi) The shape of the star in our results is rather different 
from the experimental data at $E=0$. 
In the experimental data, the ray of the star sharply extends with the 
narrow width along the $30^\circ$ direction. 
On the other hand, the ray extends with wide width in our results. 
(vii) In our results, the structure of the LDOS has cylindrical symmetry 
at a low magnetic field in the low energy case. 
However, in the experiment by Hess {\it et al.}, the clear star-shaped 
structure is observed at low fields down to 500 Gauss.

These discrepancies mean that we have to consider the other effects in 
addition to the vortex lattice effect studied here in order to reproduce 
the detailed LDOS structure of ${\rm NbSe_2}$ observed by Hess {\it et al}.  
If we consider the effect of the anisotropic $s$-wave pairing 
which has sixfold symmetrically anisotropic gap in the basal plane, 
the STM images by Hess {\it et al.} are quite well reproduced. 
We report it in detail elsewhere.\cite{Hayashi} 
It suggests that the well-studied material ${\rm NbSe_2}$ seems to  be 
one particular system which has anisotropic pairing. 
Our calculation is performed for the idealized isotropic $s$-wave 
superconductor. 
The characteristic features numerically obtained in this paper are 
expected to be well observed in other materials which are isotropic 
$s$-wave superconductors. 

To discuss the transfer of the quasiparticle bound states between 
vortex cores, we consider the LDOS structure at $E=0$. 
At high magnetic fields, the LDOS around the vortex core is connected 
each other even in the zero energy state. 
With raising a magnetic field, this connection increases.  
It means that the transfer of the quasiparticle bound state between 
vortex cores increases. 
On ${\rm NbSe_2}$, the dHvA oscillation is observed at 
magnetic fields down to 4 Tesla in the superconducting mixed state. 
We show that there is large transfer of the bound state 
between vortex cores in this high field region. 
This transfer seems to be a possible origin of the dHvA oscillation 
in superconductors. 
The estimate whether this transfer is enough for the dHvA 
oscillation or not  is a future problem.

%%%%%%%%%%%%%%%%%%%%%%%%%%%%%%%%%%%%%%%%%%%%%%%%%%%%%%%%%%%%%%%%%%%%%%
\section*{Acknowledgments}
We would like to thank H. F. Hess for fruitful discussions and 
information of their experiments. 
The authors are indebted to the Supercomputer Center of the 
Institute for Solid State Physics, University of Tokyo, and 
Kyoto University Data Processing Center, for a part of the numerical 
calculations.  
One of the authors (M. I.) is supported by a Research Fellowship of 
the Japan Society for the Promotion of Science for Young Scientists. 

%%%%%%
\newpage
%%%% figure captions %%%%%%%%%%%%%%%%%%%%%%%%%%%%%%%%%%%%%%%%%%%%%

%%%%%%
\newpage
%%%% references %%%%%%%%%%%%%%%%%%%%%%%%%%%%%%%%%%%%%%%%%%%%%%%%%%%%

\end{document}